\documentclass[pdftex,twocolumn,epjc3]{svjour3}          

\RequirePackage[T1]{fontenc}

\smartqed  

\RequirePackage{graphicx}
\RequirePackage{mathptmx}      
\RequirePackage{flushend}
\RequirePackage[numbers,sort&compress]{natbib}
\RequirePackage[colorlinks,citecolor=blue,urlcolor=blue,linkcolor=blue]{hyperref}
\usepackage{orcidlink}  
\usepackage{booktabs}
\usepackage{multirow}
\usepackage{makecell}
\usepackage{threeparttable}
\usepackage{comment}
\usepackage{graphicx}
\usepackage{dcolumn}
\usepackage{bm}
\usepackage{hyperref}
\usepackage[mathlines]{lineno}
\usepackage{amsmath}
\journalname{Eur. Phys. J. C}

\begin{document}

\title{A study of neutrinoless double electron capture in  $^{40}$Ca from the AMoRE experiment}

\author{ 
{A.~Agrawal\,\orcidlink{0000-0001-7740-5637}}
\and {V.V.~Alenkov\,\orcidlink{0009-0008-8839-0010}}
\and {P.~Aryal\,\orcidlink{0000-0003-4955-6838}}
\and {J.~Beyer\,\orcidlink{0000-0001-9343-0728}}
\and {B.~Bhandari\,\orcidlink{0009-0009-7710-6202}}
\and {R.S.~Boiko\,\orcidlink{0000-0001-7017-8793}}
\and {K.~Boonin\,\orcidlink{0000-0003-4757-7926}}
\and {O.~Buzanov\,\orcidlink{0000-0002-7532-5710}}
\and {C.R.~Byeon\,\orcidlink{0009-0002-6567-5925}}
\and {N.~Chanthima\,\orcidlink{0009-0003-7774-8367}
% ...
} \\
(AMoRE Collaboration)
}

\thankstext{corrauthor1}{e-mail: sckim@ibs.re.kr}
\thankstext{corrauthor2}{e-mail: yoomin@ibs.re.kr}
\thankstext{corrauthor3}{e-mail: bijayasharma22@gmail.com}

\date{Received: date / Accepted: date}

\maketitle

\begin{abstract}
The search for neutrinoless double electron capture ($0\nu\mathrm{2EC}$) provides a sensitive probe of lepton-number violation and the Majorana nature of neutrinos. We investigate the $0\nu\mathrm{2EC}$ decay of $^{40}$Ca using cryogenic detectors equipped with metallic magnetic calorimeters in the AMoRE-I experiment. The analysis is based on a physics dataset corresponding to a total exposure of 7.32~kg$\cdot$yr from thirteen $^{40}$Ca$^{100}$MoO$_4$ crystals. No significant excess is observed, and a lower limit on the half-life is obtained as $T^{0\nu}_{1/2} > 2.0 \times 10^{22}$~yr at 90\% confidence level. An improved sensitivity is expected for the upcoming AMoRE-II experiment. These results demonstrate the potential of CaMoO$_4$ detectors to explore rare decay processes beyond the primary $^{100}$Mo $0\nu\beta\beta$ search program.
\end{abstract}

\section{Introduction}

Electron capture (EC) is a weak-interaction process in which a bound atomic electron, typically from the innermost K or L shells, is absorbed by the nucleus, converting a proton into a neutron with the emission of an electron neutrino. The subsequent atomic relaxation produces characteristic X-rays and Auger electrons that serve as experimental signatures~\cite{bambynek1977orbital}. In some triplets, where the energy of the intermediate nucleus is higher and prevents single electron capture, two such captures can occur simultaneously, giving rise to double electron capture (2$\nu$2EC), a second-order weak process that has been directly observed in $^{124}$Xe~\cite{xenon2019,aalbers2024two,bo2025measurement}, indicated in geochemical studies of $^{130}$Ba~\cite{meshik2001weak,pujol2009xenon}, and also in $^{78}$Kr~\cite{gavrilyuk2013,ratkevich2017} by using proportional counter.
The hypothetical neutrinoless variant (0$\nu$2EC) would proceed without neutrino emission if neutrinos are Majorana particles, thus violating total lepton number conservation~\cite{bernabeu1983, vsimkovic2011, blaum2020}.
Since no neutrinos are emitted, the observable signatures of 0$\nu$2EC arise from the atomic de-excitation of the daughter atom producing X-rays and Auger electrons, as well as from the emission of at least one particle to conserve energy–momentum in the transition~\cite{karpeshin2008, doi1993neutrinoless}.
Different types of the emission of particles can be considered such as $e^-e^+$ pairs, one or two real photons, or one internal conversion electron (IC).
The probability of $0\nu\mathrm{2EC}$ decay process is extremely low, but can be enhanced significantly when the decay is nearly resonant—that is, when the energy released by the nuclear transition closely matches the sum of the binding energies of the captured electrons and the excitation energy of the daughter atom~\cite{ bernabeu1983, karpeshin2008}.
This resonant 0$\nu$2EC can shorten the half-life by many orders of magnitude, making only a handful of isotopes (such as $^{152}$Gd, $^{164}$Er, and $^{180}$W) promising candidates for current searches~\cite{bernabeu1983, vsimkovic2011, blaum2020, karpeshin2008,  Belli:2026wmw}.
The discovery of 0$\nu$2EC would not only provide direct evidence for the Majorana nature of neutrinos but also contribute to understanding the absolute neutrino mass scale and the matter–antimatter asymmetry in the Universe~\cite{bernabeu1983, vsimkovic2011, karpeshin2008, doi1993neutrinoless}, providing a complementary research avenue to neutrinoless double beta ($0\nu\beta\beta$) decay.

AMoRE is an underground experiment for the investigation of extremely rare nuclear decays with cryogenic calorimeters. The experiment makes use of scintillating crystals, operated at millikelvin temperatures, in association with metallic magnetic calorimeters for the simultaneous readout of phonon and photon signals~\cite{alenkov2019first, kmseo,amore_prl}. The dual-readout scheme enables particle identification to effectively reject backgrounds due to $\alpha$ decays~\cite{amore_prl,rodycoll}. The latest phase, AMoRE-I, has been successfully completed, demonstrating the long-term stability of the detection system for about 3 years with an upgraded setup comprising thirteen $^{40}$Ca$^{100}$MoO$_{4}$ (CMO) and five Li$_2$$^{100}$MoO$_4$ (LMO) scintillating crystals~\cite{amore_prl}.

Though the primary physics goal of AMoRE is the search for $0\nu\beta\beta$ decay of $^{100}$Mo, the CMO crystals also contain $^{40}$Ca as a naturally occurring isotope.
This makes the experiment also well-suited to search for 0$\nu$2EC of $^{40}$Ca, an especially promising candidate, considering a very high isotopic concentration in the natural isotopic admixture $\delta=96.941(156)\%$~\cite{meija2016isotopic}. The decay energy for $^{40}$Ca 0$\nu$2EC is $Q_{2\rm{EC}}=193.50(2)$ keV \cite{Wang:2021xhn}, disfavoring the possible resonant enhancement process in $^{40}$Ca isotope~\cite{blaum2020}.
An experimental investigation for 0$\nu$2EC signature of $^{40}$Ca has been previously performed by the CRESST collaboration using $\mathrm{CaWO}_4$ bolometers~\cite{cresst}.
The experiment set the half-life limit of $T_{1/2}^{0\nu} > 1.4 \times 10^{22}$~yr at 90$\%$ confidence level (CL).

We here present the experimental investigation of the 0$\nu$2EC of $^{40}$Ca based on the AMoRE-I data, using total exposure of 7.32 kg$\cdot$yr, corresponding to the $^{40}$Ca isotopic exposure of 1.43 kg$\cdot$yr.

\section{Experiment and Data analysis} 

The AMoRE-I experiment was conducted during the period between December 2020 and May 2023 at the Yangyang Underground Laboratory, located at a depth of about 700 meters of rock overburden. Detailed descriptions of the experimental setup and detector design are presented in the previous AMoRE-I publications \cite{ amore_prl, kim2022status}.
The entire AMoRE-I data have been reanalyzed--including the event trigger for lowering the energy threshold~\cite{eecs206umich}, adoption of the optimal filtering for the signal amplitude determination, corrections for signal instabilities, and the energy calibration.
While details of the analysis are provided in Ref.~\cite{my_of_paper}, we briefly describe the energy-scale reconstruction here.

The energy calibration was performed in two steps.
In the first step, amplitudes ($A$) divided by their true energy values ($E$) for four prominent peaks at 583, 911, 969, and 2615 keV of $\gamma$-rays in the $^{232}$Th source are calibrated using a linear function of energy ($A/E=p_{0}+p_{1}E$). In the second step, an additional low energy 75~keV Pb X-ray peak was included together with these $\gamma$-ray energies to refine the calibration with addition of an exponential term to the linear function, i.e. $A/E=p_{0}'+p_{1}'E+p_{2}\exp(-p_{3}E)$.
Further details on the signal analysis and calibration can be found in the Ref.~\cite{my_of_paper}.

\begin{figure}[t]%
\centering
\includegraphics[width=1\columnwidth]{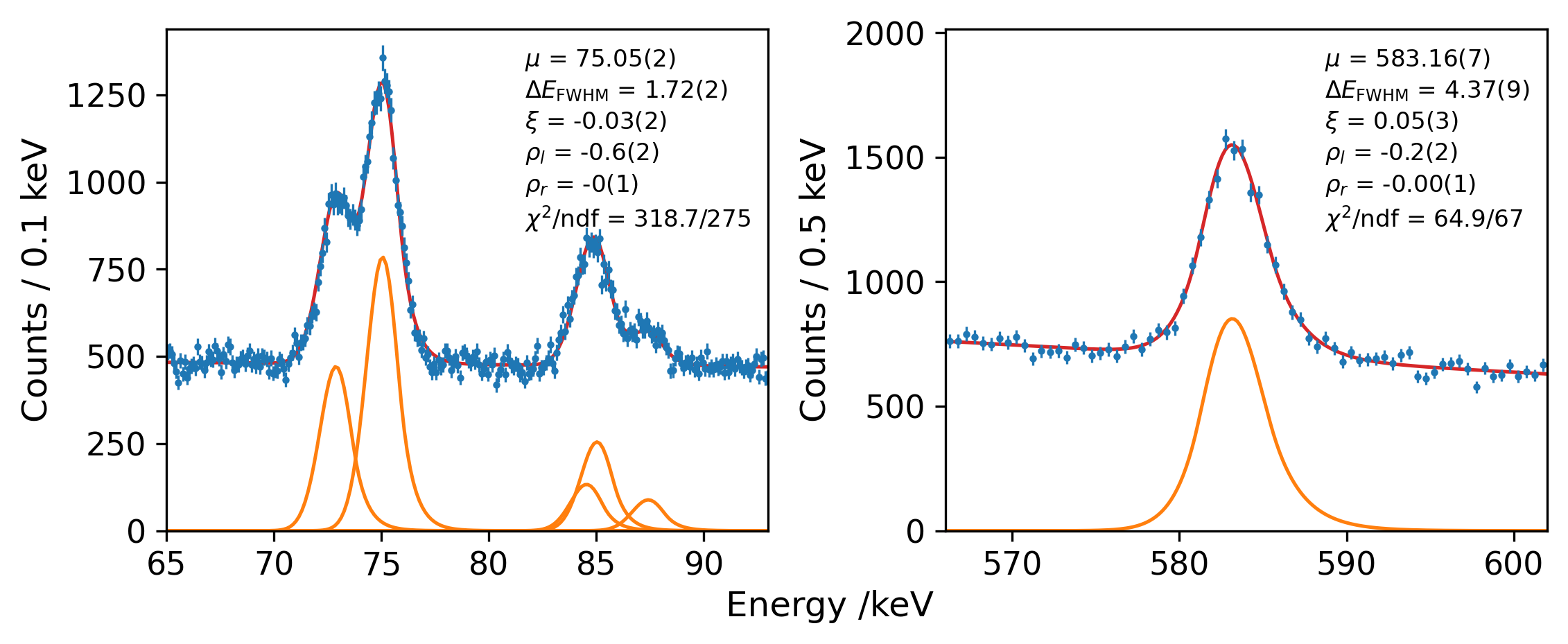}
\caption{Representative fits of the Bukin response function to calibration data from the CMO~05 detector in the APO dataset. \textit{Left panel:} Pb K-shell X-ray region containing the \(K_\alpha\) doublet near 75 keV and the \(K_\beta\) triplet around 85 keV. The individual X-ray components are fitted simultaneously with common Bukin shape parameters. \textit{Right panel:} The 583-keV \(\gamma\)-ray calibration peak. The data points are shown with statistical uncertainties; the solid red curve represents the total fit, while the individual peak components are shown separately with the orange color. The fitted peak position (\(\mu\)) and FWHM energy resolution ($\Delta E_{\mathrm{FWHM}}$) are given in keV, whereas the asymmetry parameter (\(\xi\)), and left- and right-tail parameters (\(\rho_l\)) and (\(\rho_r\)) are dimensionless. Their fitted values and uncertainties are listed in each panel.}
\label{fig1}
\end{figure}

Due to non-uniform, position-dependent detector response, energy peaks occured in most of detectors displayed asymmetric shapes with tails on the low- and high-energy sides.
Among the different tested peak fitting functions the optimal fit was provided by the Bukin function~\cite{RooBukinPdf}, which accounted for the asymmetric nature of all peaks in the calibration spectra. The detector response is characterized by the peak position ($\mu$), the full width at half maximum (FWHM) energy resolution ($\Delta E_{\mathrm{FWHM}}$), the asymmetry parameter ($\xi$), and the left-and right-tail parameters ($\rho_l$) and ($\rho_r$), respectively. Representative fits to the Pb K-shell X-ray region at 75--85~keV and to the 583-keV $\gamma$-ray calibration peak are shown in Fig.~\ref{fig1}. The FWHM energy resolution ($\Delta E_{\mathrm{FWHM}} \approx 2.355\sigma$) was modeled following previous AMoRE analyses~\cite{alenkov2019first, amore_prl} as $\Delta E_{\mathrm{FWHM}}^{2} = q_{0}+q_{1}E+q_{2}E^{2}$, to determine the value and its uncertainty at the 0$\nu$2EC region of interest (ROI). 

The response of each detector module exhibited a discontinuous shift after an unexpected power outage occurred in November 2021.
Therefore, we divided the data into two groups.  
The period from December 2020 until November 2021 is referred to as the before-power-outage (BPO) dataset, while the period from January 2022 to May 2023 is the after-power-outage (APO) dataset.
Interestingly, the energy resolution improved significantly over most of detectors after the power outage, although the underlying cause of this change remains unclear.
In addition, a detector module (CMO\,03 in Table~\ref{tab:tab1}) unresponsive during the BPO period started to work after the power outage.
The FWHM energy resolutions for all CMO crystal detectors in the two datasets are shown in Table~\ref{tab:tab1}, with most detectors exhibiting resolutions of about 2.5 to 4 keV FWHM at $0\nu\mathrm{2EC}$ ROI for APO dataset. Other shape parameters like asymmetry, left-tail and right-tail, which are obtained by fitting the calibration peaks, showed weak dependence on energy, and their uncertainties were interpolated using a linear or exponential function based on detector's characteristics to determine the signal shape at ROI.

\begin{table*}[t] 
\centering
\caption{Summary of mass-time exposure ($m{\cdot}t$), FWHM energy resolution ($\Delta E_{\mathrm{FWHM}}$), and background index ($b$) in the ROI for the AMoRE-I CMO crystal detectors in the BPO and APO datasets.
Uncertainty of $\Delta E_{\mathrm{FWHM}}$ is typically at a few-percent level.
The higher background in the BPO dataset is attributed to temporary aluminum plates in the shielding supporting structure, later found to be contaminated with $^{228}$Ra and removed~\cite{amore_prl}.
Detector modules indexes are for the CMO detectors only, listed with their masses in Ref.~\cite{kim2022status}.}
\label{tab:tab1}
\centering
\setlength{\tabcolsep}{6pt}

\begin{tabular*}{\textwidth}{@{\extracolsep{\fill}}ccccccc@{}}
\toprule
\multirow{2}{*}[-0.2em]{\makecell{Detector module\\index}} 
& \multicolumn{2}{c}{$m\cdot t$ ($10^{-3}\;$kg${\cdot}$yr)} 
& \multicolumn{2}{c}{$\Delta E_{\mathrm{FWHM}}$ (keV)} 
& \multicolumn{2}{c}{$b$ (count/keV/kg/day)} \\
\cmidrule(lr){2-3}\cmidrule(lr){4-5}\cmidrule(lr){6-7}
& BPO & APO & BPO & APO & BPO & APO \\
\midrule
01 & 232 & 384 & 5.8 & 4.5 & 7.4 & 6.1 \\
02 & 232 & 356 & 2.9 & 2.5 & 7.9 & 5.4 \\
03 & --  & 342 & --  & 5.3 & --  & 5.7 \\
04 & 235 & 362 & 2.9 & 2.6 & 8.4 & 5.7 \\
05\rlap{$^{*}$} & 169 & 258 & 3.4 & 2.6 & 59.8 & 54.7 \\
06 & 232 & 302 & 5.0 & 6.5 & 8.5 & 5.7 \\
07 & 281 & 428 & 3.3 & 3.0 & 9.1 & 7.5 \\
08 & 230 & 359 & 3.8 & 3.3 & 8.5 & 5.6 \\
09 & 233 & 356 & 3.2 & 2.7 & 8.4 & 6.2 \\
10 & 126 & 197 & 3.3 & 2.9 & 10.2 & 6.5 \\
11 & 311 & 473 & 6.0 & 3.6 & 7.1 & 5.3 \\
12 & 246 & 375 & 4.2 & 3.2 & 8.8 & 6.1 \\
13 & 236 & 360 & 4.3 & 3.0 & 8.6 & 6.2 \\
\midrule
\makecell{Total /\\Average} 
& \multicolumn{2}{c}{$7\,316$} 
& \multicolumn{2}{c}{3.4} 
& \multicolumn{2}{c}{9.7} \\
\bottomrule
\end{tabular*}

\vspace{1mm}
\begin{minipage}{0.92\linewidth}
\footnotesize
$^{*}$ CMO\,05 shows an unusually high background compared to the other detectors due to contamination from U-series and Ac-series isotopes.
\end{minipage}

\end{table*}

\begin{figure}[t]%
\centering
\includegraphics[width=1\columnwidth]{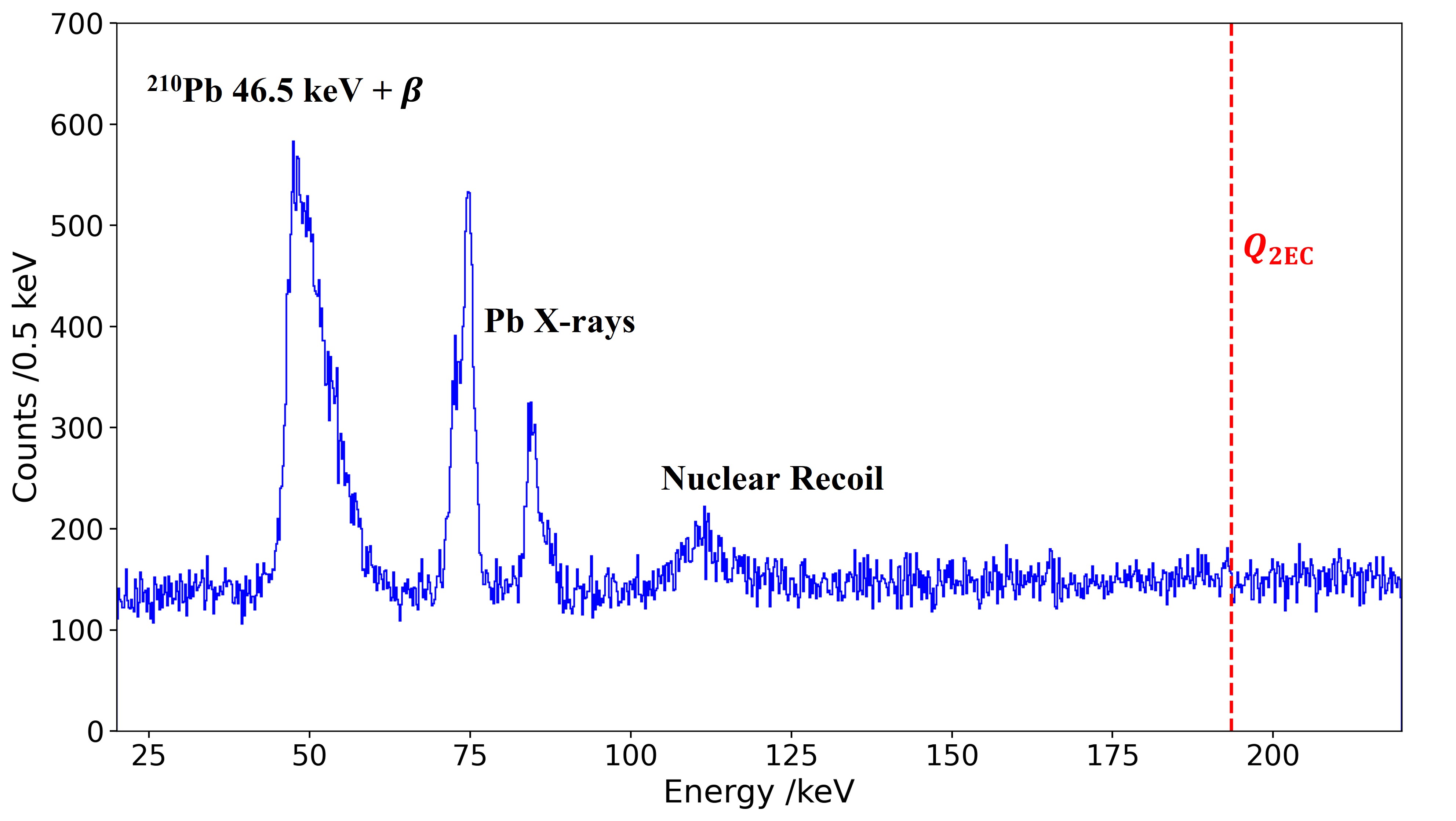}
\caption{Background spectrum of the CMO\,02 APO dataset. 
The peak around 50~keV originates from the beta decay of $^{210}$Pb, which consists of 46.5 keV IC/Auger electrons and $\gamma$/X-ray together with about 4~keV mean energy deposition from $\beta$ electrons~\cite{nudat, adhikari2021background}. 
From left to right, distinct features are observed: the Pb $K_{\alpha}$ doublet near 75~keV, the $K_{\beta}$ triplet around 85~keV~\cite{thompson2001x}, and a peak at $\sim$103~keV associated with the nuclear recoil of $^{206}$Pb following the $^{210}$Po $\alpha$ decay with 5304 keV emission~\cite{nudat}. The dashed red line represents the $Q_{2\mathrm{EC}}$ value of $0\nu\mathrm{2EC}$ decay process.}
\label{fig2}
\end{figure}

Basic data-quality conditions were applied to eliminate non-physical events such as heater pulses and SQUID reset signals~\cite{my_of_paper}.
To suppress backgrounds from cosmic-ray interactions and multi-site energy depositions, coincident events in other detector channels or in the external muon veto system were rejected.
The muon detector registered about ten thousand muon candidates per day~\cite{amore_prl}, and the resulting coincidence cut within a 3 ms window reduced the efficiency by less than 0.1\%, with negligible associated uncertainty. 
In addition, a single-hit requirement was imposed: only events with no additional signal at other detector module within a 3 ms time window were retained.
Given the typical trigger rate of $\sim$1 Hz during physics runs, this selection achieved an event selection efficiency of approximately 99.8\%.
The typical trigger threshold was around 20~keV for most detectors, well below the ROI near 193.5~keV for $^{40}$Ca $0\nu$2EC.
Figure~\ref{fig2} shows an energy spectrum of the detector CMO\,02 for APO dataset where we can observe low energetic $\gamma$ line/IC, X-rays, and nuclear recoil signature. 

The $0\nu$2EC process in $^{40}$Ca can proceed through distinct atomic capture processes depending on the shells of the captured electrons.
The dominant capture probabilities are $f_{K\!K}\simeq85\%$, $f_{L\!L}\simeq1\%$ ~\cite{cresst}, with the remaining $f_{K\!L}\simeq14\%$ inferred to complete the total capture probability.

For a ground- to ground-state transition, the emission of a single real photon is forbidden in the KK capture channel by angular momentum conservation~\cite{doi1993neutrinoless}.
While double-photon emission is kinematically allowed, it remains highly suppressed as a second-order electromagnetic process, scaling with an additional factor of the fine-structure constant and subject to a reduced phase-space volume.
Consequently, the transition proceeds predominantly via IC, in which a virtual photon is absorbed by an atomic electron, resulting in the emission of an IC electron.
In contrast, both single real photon emission and IC are allowed for KL capture.

The full-energy absorption efficiency was evaluated using \textsc{GEANT4} Monte Carlo simulations~\cite{AGOSTINELLI2003250}, incorporating the realistic detector geometry and material properties of the CMO crystals.
For each decay channel, the total available energy is given by the $Q$-value of transition and is redistributed according to the binding energies of the captured electrons.
In all channels, atomic vacancies created by electron capture lead to the emission of X-rays and Auger electrons during atomic relaxation.
Since these particles deposit their energy locally in the crystal, the total observable energy corresponds to the $Q$ value of the decay, resulting in a mono-energetic peak.

Three representative channels were simulated: (i) KK capture with an IC electron, (ii) KL capture with a real photon, and (iii) KL capture with an IC electron.
For the IC channels, most of the emitted electrons deposit their full energy in the detector, resulting in the detection efficiencies close to unity, i.e., $\varepsilon^{\mathrm{IC}}\approx1$.
In contrast, the KL photon channel shows a lower full-energy absorption efficiency, $\varepsilon_{\rm{K\!L}}^{\gamma} \approx 0.47\text{--}0.57$, depending on the crystal mass (196–473 g).
This reflects the reduced photon interaction probability in CMO crystals (density 4.34 g/cm$^{3}$) compared with CaWO$_4$ (6.06 g/cm$^{3}$) crystals used in CRESST-II~\cite{cresst}

The containment detection efficiency ($\varepsilon_{\mathrm{cont}}$) is obtained by weighting the individual detection efficiencies by the corresponding capture probabilities,
\begin{equation}
\varepsilon_{\mathrm{cont}} =
f_{\mathrm{K\!K}}\,\varepsilon_{\mathrm{K\!K}}^{\mathrm{IC}}
+
f_{\mathrm{K\!L}}
\left(
r_{\gamma}\,\varepsilon_{\mathrm{K\!L}}^{\gamma}
+
r_{\mathrm{IC}}\,\varepsilon_{\mathrm{K\!L}}^{\mathrm{IC}}
\right),
\label{eq:branching_ratio}
\end{equation}
where $r_{\gamma}$ and $r_{\mathrm{IC}}\;(=\!1\!-\!r_{\gamma})$ denote the branching fractions for photon emission and IC in the KL capture channel, respectively.
We assumed $r_{\gamma}=1$ to conservatively determine half-life, as the exact ratio between $r_{\gamma}$ and $r_{\mathrm{IC}}$ remains undetermined. Under this assumption, Eq.~\ref{eq:branching_ratio} reduces to 
\begin{equation}
\varepsilon_{\mathrm{cont}} =
f_{\mathrm{K\!K}}\,\varepsilon_{\mathrm{K\!K}}^{\mathrm{IC}}
+
f_{\mathrm{K\!L}}\,r_{\gamma}\,\varepsilon_{\mathrm{K\!L}}^{\gamma}
\label{eq:branching_ratio2}
\end{equation}

The resulting overall containment detection efficiency for the CMO crystals is found to lie in the range $\varepsilon_{\mathrm{cont}}= 0.91\!-\!0.93$.

The total detection efficiency ($\varepsilon$) for the $0\nu\mathrm{2EC}$ signature is obtained by combining the containment efficiency with the anti-coincidence survival efficiency and the trigger efficiency. The trigger efficiency at the ROI is unity for all detectors except for 0.93 for the noisiest CMO\,01 detector.
Taking these factors into account for each detector, the resulting detection efficiency is estimated to be $\epsilon \approx 0.86\text{--}0.92$.

\section{Limit on the $0\nu2\mathrm{EC}$ half-life of $^{40}\mathrm{Ca}$}

The search for the $^{40}$Ca $0\nu2$EC decay was performed by fitting the energy spectra in a $\pm~12\sigma$ energy window around the expected signal peak at $Q_{2\rm{EC}} = 193.5$~keV. The energy spectrum of the $i$'th dataset is modeled as the sum of background components ($B_{i}$) and an expected 0$\nu$2EC signal ($S_{i}$).
$B_{i}$ consist of a slowly varying part described by a linear function of energy ($B^{\mathrm{lin}}_{i}$), superimposed with a small peak at 186~keV ($B^{\mathrm{peak}}_{i}$) which is primarily attributed to the decay of $^{226}$Ra with a possible contribution from the decay of $^{235}$U~\cite{alenkov2022alpha}.
The expected signal is expressed as:

\begin{equation}
    S_{i}(\varGamma,E;\boldsymbol{\theta}_{i}) = \varepsilon_{i}\varGamma\frac{\eta\,m_{i}\,N_{\!A}}{M_{\mathrm{CMO}}}t_{i}\,P(E;\boldsymbol{\theta}_{i}),
\end{equation}
where $\varepsilon_{i}$ denote the total detection efficiency, $\varGamma$ is the $0\nu2$EC decay rate, $\eta$ is the abundance of $^{40}$Ca in the CMO crystal. The calcium used in the AMoRE crystals was depleted in $^{48}$Ca to suppress the $2\nu\beta\beta$ background from $^{48}$Ca, resulting in an enhanced $^{40}$Ca abundance of $\eta\!\simeq\!99.95\%$~\cite{Karki:2018rhc}. $N_A$ is Avogadro's number, $m_i$ is the crystal mass, $t_i$ is the live time, and $M_{\mathrm{CMO}}$ is the molar mass of CMO compound.
The function $P(E;\boldsymbol{\theta}_{i})$ denotes the Bukin probability density function (p.d.f.) describing the detector response~\cite{RooBukinPdf}, where $\boldsymbol{\theta}_{i}$ is the set of the function parameters consisting of the peak location and the shape parameters such as the resolution, asymmetry, left-tail and right-tail.
$B_{i}^{\mathsf{peak}}$ is also represented using the Bukin function sharing same values of the shape parameters with $S_{i}$, because the locations of the two peaks are very close.
The shape parameters and the peak locations were subject to Gaussian constraints, whose widths ($\varDelta\boldsymbol{\theta}_{i}$) were derived in the energy calibration process.

We then construct a binned likelihood $\mathcal{L}$ combining the above probability density functions and constraints:
\begin{equation}
    \mathcal{L}=\prod_{i}\mathcal{L}_{i}(\varGamma, \mu_{i}; \boldsymbol{\theta}_{i})=\prod_{i}\!\left[ \prod_{j}\frac{\mu_{ij}^{n_{ij}}e\!^{-\mu_{ij}}}{n_{ij}!}\right]e\!^{-\frac{(\boldsymbol{\theta}'_{i}-\langle\boldsymbol{\theta}'_{i}\rangle)^{2}}{2(\varDelta\boldsymbol{\theta}'_{i})^{2}}},
    \label{eq:likelihood}
\end{equation}
where $n_{ij}$ is the observed event number in the $j$'th energy bin of the $i$'th dataset, $\mu_{ij}$ is the expected event number consisting of the total number of events and the model p.d.f.~with the energy bin width $\varDelta E = 0.1$~keV:
\begin{equation}
\mu_{ij}=\mu_{i}\left[ f_{i}B_{i}(E_{j})+(1\!-\!f_{i})S_{i}(E_{j})\right]\!\cdot\!\varDelta E,
\end{equation} where $f_i$ and $1-f_i$ denote the background and signal fractions, respectively
and the last exponent term accounts for the Gaussian constraints for systematic uncertainties regarding all efficiencies and peak function parameters, with the mean values $\langle\boldsymbol{\theta}'_{i}\rangle$ and widths $\varDelta\boldsymbol{\theta}'_{i}$.
The fitting was performed using the {\sc iminuit} software~\cite{dembinski_2025_17565861}, to calculate the minimized negative logarithms of $\mathcal{L}$ in Eq.~\ref{eq:likelihood} along the $\varGamma$ values from 0 to the sensitive range at around $10^{-21}$~yr$^{-1}$.

\section{Results and Discussions}

\begin{figure}[t]%
\centering
\includegraphics[width=0.98\columnwidth]{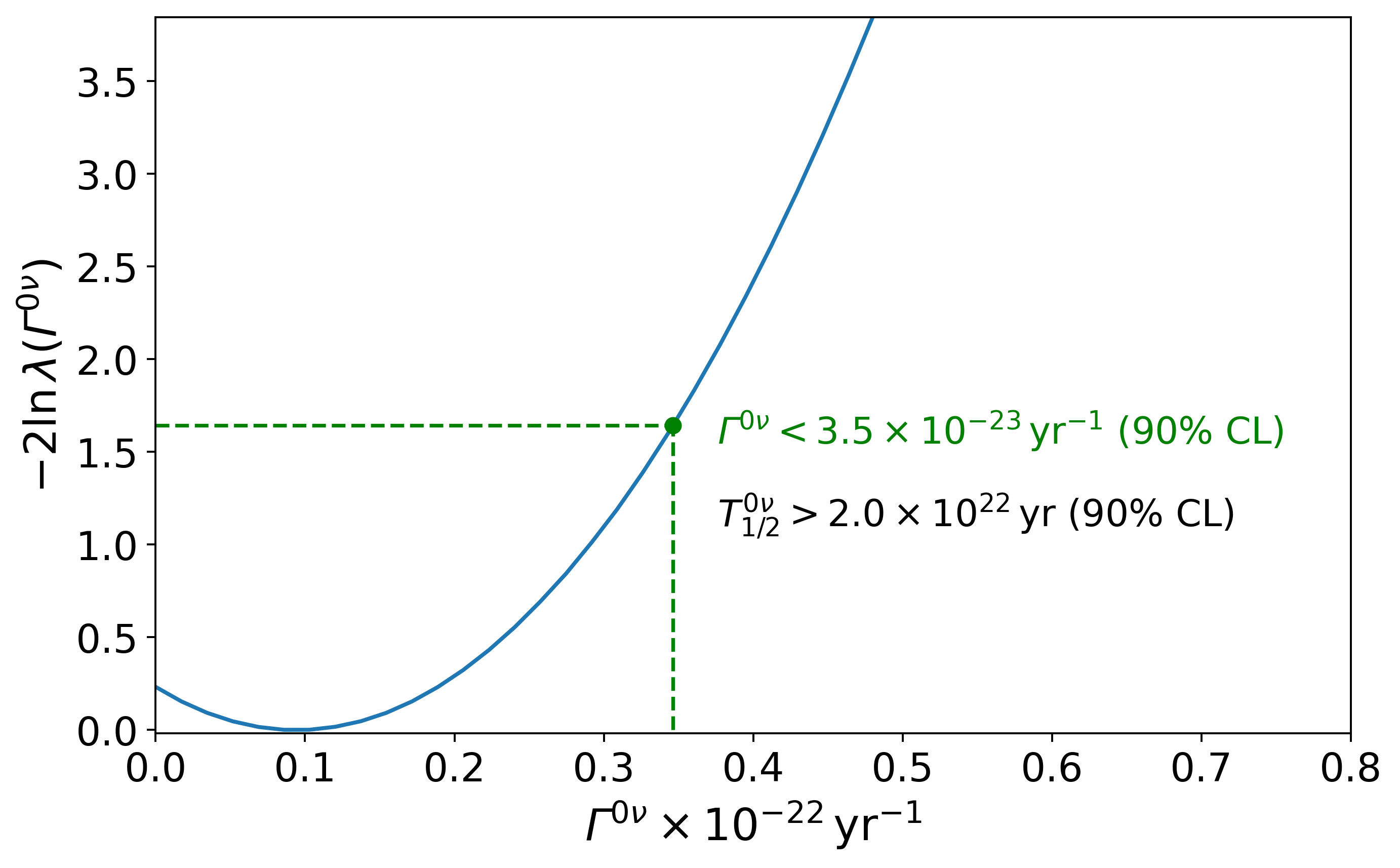}
\caption{Profile-likelihood-ratio test statistics, $q(\Gamma) = -2\ln\lambda(\varGamma^{0\nu})$, as a function of $0\nu\mathrm{2EC}$ decay rate. The horizontal green dashed line indicates the one-sided 90$\%$ CL criterion, while the vertical green dashed line shows the corresponding upper limit of $\varGamma$.}
\label{fig3}
\end{figure}

\begin{figure}[t]%
\centering
\includegraphics[width=0.99\columnwidth]{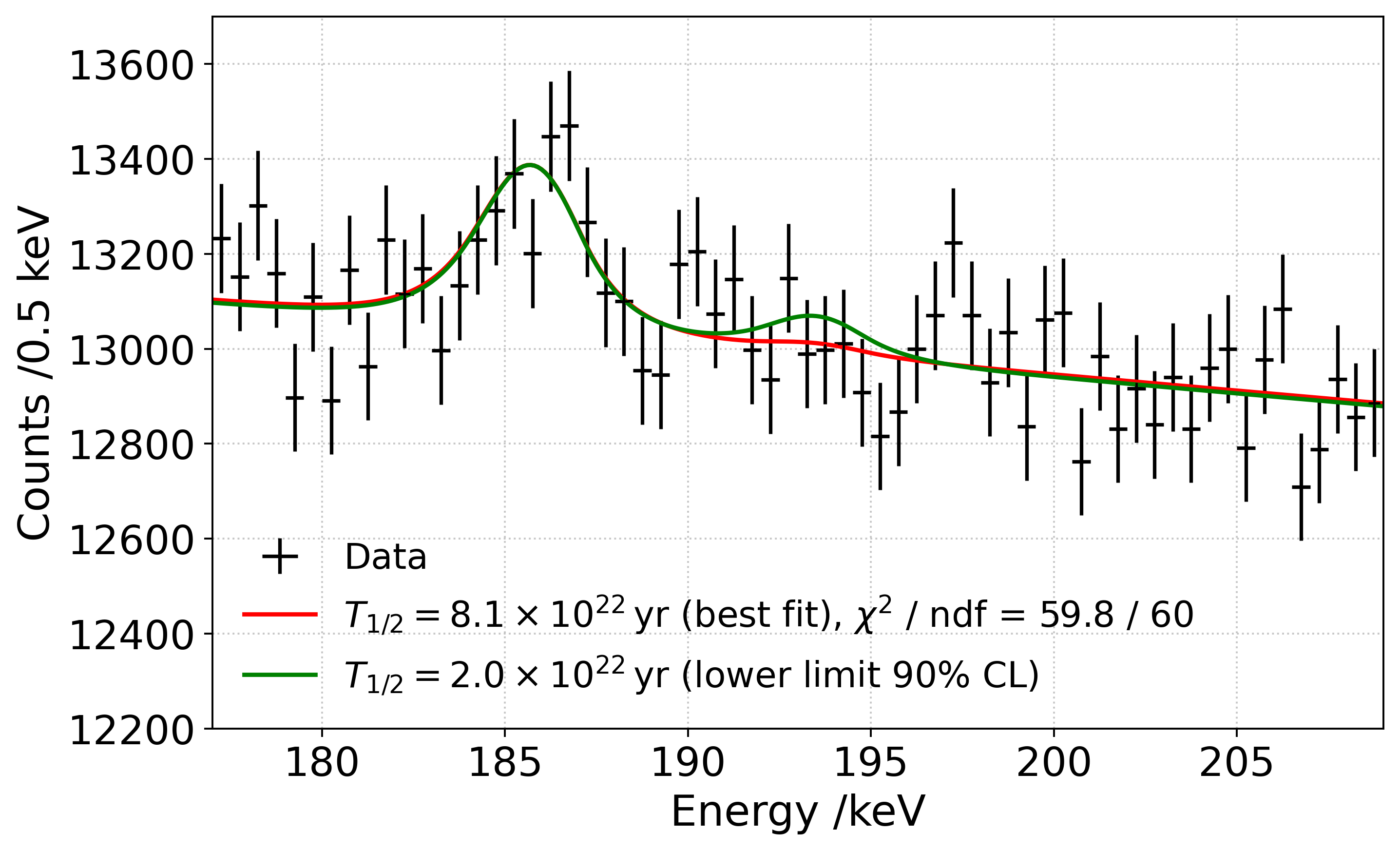}
\caption{A combined energy spectrum of all 13 CMO detectors with best fit (red curve) of the signal-plus-background model. The 90$\%$ CL limit is also shown by green curve. The peak around 186 keV is attributed to unresolved $\gamma$-ray background contributions, as discussed in the text.}
\label{fig4}
\end{figure}

Figure~\ref{fig3} shows the profile-likelihood-ratio test statistic,
$q(\varGamma)=-2\ln\lambda(\varGamma)$, as a function of the
$0\nu\mathrm{2EC}$ decay rate $\varGamma$. The profile-likeklihood ratio is defined as 
$\lambda(\varGamma)=\mathcal{L}(\varGamma,\hat{\hat{\boldsymbol{\theta}}}_{\varGamma})/
\mathcal{L}(\hat{\varGamma},\hat{\boldsymbol{\theta}})$, where $\hat{\varGamma}$ and $\hat{\boldsymbol{\theta}}$ denote
the global best-fit values of the decay rate and nuisance parameters, respectively, while
$\hat{\hat{\boldsymbol{\theta}}}_{\varGamma}$ represents the best-fit nuisance parameters for a fixed value of $\varGamma$. The minimum occurs at
$\hat{\varGamma}\sim8.6\times10^{-24}$~yr$^{-1}$. This non-zero value of $\hat{\varGamma}$ does not signal evidence for a $0\nu\mathrm{2EC}$ signal; rather, it reflects statistical fluctuations in the background within the ROI. Figure~\ref{fig4} displays the best-fit including signal and background model for combined datasets.
The upper limit on $\varGamma$, or equivalently the lower limit on the $0\nu\mathrm{2EC}$ half-life ($T_{1/2}^{0\nu}$), is determined using the
profile-likelihood-ratio approach using critical value $q = 1.64$. We obtain $\varGamma<3.5\times10^{-23}$~yr$^{-1}$
at 90\% CL, corresponding to a lower half-life limit of $T_{1/2}^{0\nu}>2.0\times10^{22}$~yr.
This presents stronger constraint for the half-life compared to earlier study reported by CRESST-II experiment using CaWO$_{4}$ crystal \cite{cresst}.

The experimental sensitivity was estimated using pseudo-experiments under the background-only hypothesis.
Thousands of pseudo-datasets were generated assuming the background model without a signal contribution.
Each pseudo-dataset was then fitted with the same signal-plus-background model used in the data analysis, and the upper limit of the decay rate $\varGamma$ at 90\% CL was obtained using the same profile-likelihood-ratio procedure used for the observed limit.
The median~$\pm$~34\% exclusion sensitivity obtained from these pseudo-experiments corresponds to a half-life limit of $T_{1/2}^{0\nu} = 2.5^{+2.6}_{-1.1} \times 10^{22}\ \mathrm{yr}.$

As the collaboration prepares for the AMoRE-II experiment, which will soon become operational employing 90 kg of $^{100}$Mo, most of the detector array will consist of LMO crystals.
However, at least 13 CMO crystals previously used in the AMoRE-I experiment will also be installed in the AMoRE-II setup. 
Considering dedicated studies for the background reduction, optimized shield configuration~\cite{agrawal2024radioassay, prbkg}, and enhanced detector performance~\cite{rodycoll, kim2022optimization} for AMoRE-II, significantly lower background level is expected with the new experiment even at the lower energy relevant to $0\nu\mathrm{2EC}$ in $^{40}$Ca.
Based on the simulated background level at the ROI, we projected the exclusion sensitivity of AMoRE-II on the $0\nu\mathrm{2EC}$ decay of $^{40}$Ca.
The estimate assumes a total CMO detector mass of 4.6~kg (similar to AMoRE-I) and a live time of 5~years, with the energy resolution and detection efficiency obtained in the AMoRE-I experiment.
The resulting median~$\pm$~34\% exclusion sensitivity is $T_{1/2}^{0\nu} \sim 1^{+1}_{-0.5}\!\times \!10^{23}$~yr at 90\% CL, representing an improvement of about a factor of five compared to the current AMoRE-I limit.
This improvement arises primarily from the expected fivefold reduction in background levels together with an approximately threefold increase in exposure. 

\section{Summary and Conclusions}

A search for the $0\nu\mathrm{2EC}$ decay of $^{40}$Ca was performed using the AMoRE-I background dataset, corresponding to a total exposure of 7.32~kg$\cdot$yr and an isotopic exposure of 1.43~kg$\cdot$yr in $^{40}$Ca.
No excess consistent with the signal was observed, and a lower limit on the half-life was set as $T_{1/2}^{0\nu} > 2.0 \times 10^{22}~\mathrm{yr}$ at 90\% CL.
This result slightly improves the previous limit reported by the CRESST collaboration.

The present analysis demonstrates that the AMoRE experiment can probe rare processes beyond its primary search for the $0\nu\beta\beta$ decay of $^{100}$Mo. 
In particular, CMO crystals provide a promising experimental platform for investigating the $0\nu\mathrm{2EC}$ decay of $^{40}$Ca with competitive sensitivity.
In preparation for the AMoRE-II experiment, a sensitivity study was performed assuming a CMO detector mass of 4.6~kg, a live time of 5~years, and the background level obtained from simulations of the improved experimental configuration.
Using the energy resolution and detection efficiency estimated in AMoRE-I, the expected median exclusion sensitivity was estimated to be $T_{1/2}^{0\nu} \sim 1\!\times\!10^{23}~\mathrm{yr}$ at 90\% CL, representing an improvement of about a factor five compared to the present AMoRE-I limit.

Furthermore, the current AMoRE-I dataset does not reach the sufficiently low energy threshold required to search for the allowed two-neutrino double K-capture ($2\nu\mathrm{2K}$) process in $^{40}$Ca, whose expected signature lies at 6.4~keV~\cite{thompson2001x} and which has not yet been observed.
However, with the improved detector performance, enhanced noise suppression, and optimized trigger algorithms anticipated for the AMoRE-II experiment, searches for both the $2\nu\mathrm{2K}$ and $0\nu\mathrm{2EC}$ processes in $^{40}$Ca are expected to become feasible with substantially improved sensitivity.

These results highlight the broader physics potential of the AMoRE project, extending from few-keV rare-event searches to MeV-scale probes of physics beyond the Standard Model.

\begin{acknowledgements}

The authors thank Prof. M.\,I. Krivoruchenko (Laboratory of Elementary Particles Theory, Kurchatov Complex of Theoretical and Experimental Physics, National Research Centre ``Kurchatov Institute", Moscow, Russia) for valuable discussions on the particle emission mechanisms in the $0\nu\mathrm{2EC}$ process of $^{40}$Ca.

This research was supported by the Institute for Basic Science (IBS) under Grants No.~IBS-R016-D1 and No.~IBS-R016-A2. Additional support was provided by the National Research Foundation of Korea (NRF) under Grants No.~NRF-2021R1I1A3041453, No.~NRF-2021R1A6A1A03043957, RS-2025-20802968, and by the National Research Facilities \& Equipment Center (NFEC) of Korea under Grant No.~2019R1A6C1010027.

We also acknowledge support from the Ministry of Science and Higher Education of the Russian Federation (Project No.~N121031700314-5) and the MEPhI Program Priority 2030.

We thank the Korea Hydro and Nuclear Power (KHNP) Company for providing underground laboratory space at Yangyang and the IBS Research Solution Center (RSC) for providing high-performance computing resources.

These acknowledgments are not to be interpreted as an endorsement of any statement made by the authors by any of the affiliated institutes, funding agencies, governments, or their representatives.
\end{acknowledgements}


\begin{thebibliography}{9}

\bibitem{bambynek1977orbital}
W. Bambynek et al.,
Rev. Mod. Phys. \textbf{49}, 77--221 (1977).
https://doi.org/10.1103/RevModPhys.49.77


\bibitem{xenon2019}
E. Aprile et al. (XENON Collaboration),
Nature \textbf{568}, 532--535 (2019).
https://doi.org/10.1038/s41586-019-1124-4


\bibitem{aalbers2024two}
J. Aalbers et al. (LZ Collaboration),
J. Phys. G: Nucl. Part. Phys. \textbf{52}, 015103 (2025);
Erratum: J. Phys. G: Nucl. Part. Phys. \textbf{53}, 059601 (2026).
https://doi.org/10.1088/1361-6471/ad9039


\bibitem{bo2025measurement}
Z. Bo et al. (PandaX-4T Collaboration),
JHEP \textbf{05}, 119 (2025).
https://doi.org/10.1007/JHEP05(2025)119


\bibitem{meshik2001weak}
A.P. Meshik et al.,
Phys. Rev. C \textbf{64}, 035205 (2001).
https://doi.org/10.1103/PhysRevC.64.035205

\bibitem{pujol2009xenon}
M. Pujol et al., Geochimica et
Cosmochimica Acta \textbf{73}, 6834--6846 (2009).
https://doi.org/10.1016/j.gca.2009.08.002

\bibitem{gavrilyuk2013}
Yu.M. Gavrilyuk et al.,
Phys. Rev. C \textbf{87}, 035501 (2013).
https://doi.org/10.1103/PhysRevC.87.035501

\bibitem{ratkevich2017}
S.S. Ratkevich et al.,
Phys. Rev. C \textbf{96}, 065502 (2017).
https://doi.org/10.1103/PhysRevC.96.065502

\bibitem{bernabeu1983}
J. Bernab\'{e}u, A. De R\'{u}jula, C. Jarlskog,
Nucl. Phys. B \textbf{223}, 15--28 (1983).
https://doi.org/10.1016/0550-3213(83)90089-5

\bibitem{vsimkovic2011}
F. \v{S}imkovic, M.I. Krivoruchenko, A. Faessler,
Prog. Part. Nucl. Phys. \textbf{66}, 446--451 (2011).
https://doi.org/10.1016/j.ppnp.2011.01.049


\bibitem{blaum2020}
K. Blaum et al.,
Rev. Mod. Phys. \textbf{92}, 045007 (2020).
https://doi.org/10.1103/RevModPhys.92.045007

\bibitem{karpeshin2008}
F.F. Karpeshin,
Phys. Part. Nucl. Lett. \textbf{5}, 379--382 (2008).
https://doi.org/10.1134/S1547477108040080


\bibitem{doi1993neutrinoless}
M. Doi, T. Kotani,
Prog. Theor. Phys. \textbf{89}, 139--159 (1993).
https://doi.org/10.1143/ptp/89.1.139

\bibitem{Belli:2026wmw}
P. Belli et al.,
J. Phys. G: Nucl. Part. Phys. \textbf{53}, 045101 (2026).
https://doi.org/10.1088/1361-6471/ae4d66

\bibitem{alenkov2019first}
V. Alenkov et al.,
Eur. Phys. J. C \textbf{79}, 791 (2019).
https://doi.org/10.1140/epjc/s10052-019-7279-1

\bibitem{kmseo}
A. Agrawal et al. (AMoRE Collaboration),
Astropart. Phys. \textbf{162}, 102991 (2024).
https://doi.org/10.1016/j.astropartphys.2024.102991

\bibitem{amore_prl}
A. Agrawal et al. (AMoRE Collaboration),
Phys. Rev. Lett. \textbf{134}, 082501 (2025).
https://doi.org/10.1103/PhysRevLett.134.082501

\bibitem{rodycoll}
A. Agrawal et al. (AMoRE Collaboration),
Eur. Phys. J. C \textbf{85}, 172 (2025).
https://doi.org/10.1140/epjc/s10052-024-13498-8

\bibitem{meija2016isotopic}
J. Meija et al.,
Pure Appl. Chem. \textbf{88}, 293--306 (2016).
https://doi.org/10.1515/pac-2015-0503

\bibitem{Wang:2021xhn}
M. Wang et al.,
Chin. Phys. C \textbf{45}, 030003 (2021).
https://doi.org/10.1088/1674-1137/abddaf

\bibitem{cresst}
G. Angloher et al.,
J. Phys. G: Nucl. Part. Phys. \textbf{43}, 095202 (2016).
https://doi.org/10.1088/0954-3899/43/9/095202


\bibitem{kim2022status}
H.B. Kim et al.,
J. Low Temp. Phys. \textbf{209}, 962--970 (2022).
https://doi.org/10.1007/s10909-022-02880-z

\bibitem{eecs206umich}
M.A. Bartsch, D.L. Neuhoff, G.H. Wakefield,
University of Michigan EECS 206 Laboratory Manual (2003).
\url{https://www.eecs.umich.edu/courses/eecs206/public/lab/}

\bibitem{my_of_paper}
A. Agrawal et al. (AMoRE Collaboration),
The Development of Analysis Framework for AMoRE Experiment,
in preparation.

\bibitem{RooBukinPdf}
RooBukinPdf Class Reference.
\url{https://root.cern.ch/doc/master/classRooBukinPdf.html}
Accessed 16 September 2025.


\bibitem{nudat}
National Nuclear Data Center,
NuDat 3.0 Database, Brookhaven National Laboratory.
\url{https://www.nndc.bnl.gov/nudat3/}
Accessed March 2026.

\bibitem{adhikari2021background}
G. Adhikari et al. (COSINE-100 Collaboration),
Eur. Phys. J. C \textbf{81}, 837 (2021).
https://doi.org/10.1140/epjc/s10052-021-09564-0

\bibitem{thompson2001x}
A.C. Thompson, D. Vaughan (eds.),
X-Ray Data Booklet, 2nd edn.
Lawrence Berkeley National Laboratory, Berkeley (2001).
\url{https://xdb.lbl.gov/}

\bibitem{AGOSTINELLI2003250}
S. Agostinelli et al.,
Nucl. Instrum. Methods Phys. Res. A \textbf{506}, 250--303 (2003).
https://doi.org/10.1016/S0168-9002(03)01368-8

\bibitem{alenkov2022alpha}
V. Alenkov et al. (AMoRE Collaboration),
Eur. Phys. J. C \textbf{82}, 1140 (2022).
https://doi.org/10.1140/epjc/s10052-022-11104-3


\bibitem{Karki:2018rhc}
S.~Karki et al., Nucl. Instrum. Methods Phys. Res. A \textbf{877}, 328--330 (2018). https://doi.org/10.1016/j.nima.2017.10.007

\bibitem{dembinski_2025_17565861}
H. Dembinski et al., scikit-hep/iminuit, Zenodo (2025). https://doi.org/10.5281/zenodo.17565861

\bibitem{agrawal2024radioassay}
A. Agrawal et al. (AMoRE Collaboration), Front. Phys. \textbf{12}, 1362209 (2024).
https://doi.org/10.3389/fphy.2024.1362209

\bibitem{prbkg}
A. Agrawal et al. (AMoRE Collaboration), Eur. Phys. J. C \textbf{85}, 9 (2025).
https://doi.org/10.1140/epjc/s10052-024-13516-9

\bibitem{kim2022optimization}
W.T. Kim et al., JINST \textbf{17}, P07034 (2022).
https://doi.org/10.1088/1748-0221/17/07/P07034

\end{thebibliography}
\end{document}